\documentclass[twocolumn]{aastex631}
\usepackage[utf8]{inputenc}
\usepackage{multirow}
\usepackage[numbib]{tocbibind}
\usepackage{amsmath,amsthm,amssymb}
\usepackage{longtable}
\usepackage[T1]{fontenc}
\usepackage{threeparttable}

\begin{document}

\title{Functionality of Ice Line Latitudinal EBM Tenacity (FILLET). Protocol Version 1.1}
\author[0000-0001-6487-5445]{Rory Barnes}
\affil{Astronomy Department, University of Washington, Seattle, WA, USA 98105-1580}
\email{rory@astro.washington.edu}
\author[0000-0001-9423-8121]{Russell Deitrick}
\affil{School of Earth and Ocean Sciences, University of Victoria, Victoria, British Columbia, Canada}

\author[0000-0003-4346-2611]{Jacob Haqq-Misra}
\affil{Blue Marble Space Institute of Science, Seattle, WA, USA}
\author[0000-0002-5826-1540]{Shintaro Kadoya}
\affil{Japan Agency for Marine-Earth Science and Technology, X-star, Kanagawa, Japan}
\author[0000-0001-7553-8444]{Ramses Ramirez}
\affil{University of Central Florida, Department of Physics, Planetary Sciences Group, Orlando, Fl. 32816}
\author[0000-0002-7744-5804]{Paolo Simonetti}
\affiliation{University of Trieste, Dep. of Physics, Via G.~B. Tiepolo 11, I-34143 Trieste, Italy}
\affiliation{INAF Trieste Astronomical Observatory, Via G.~B. Tiepolo 11, I-34143 Trieste, Italy}
\author[0000-0002-5638-4344]{Vidya Venkatesan}
\affil{Department of Physics and Astronomy, University of California, Irvine, 4129 Frederick Reines Hall, Irvine, CA 92697-4575, USA }
\author[0000-0002-5967-9631]{Thomas J. Fauchez}
\affiliation{NASA Goddard Space Flight Center,
8800 Greenbelt Road
Greenbelt, MD 20771, USA}
\affiliation{Integrated Space Science and Technology Institute, Department of Physics, American University, Washington DC}
\affiliation{NASA GSFC Sellers Exoplanet Environments Collaboration}

\date{14 Nov 2025}

\begin{abstract}
The Functionality of Ice Line Latitudinal EBM Tenacity (FILLET) project is a CUISINES exoplanet model intercomparison project (exo-MIP) that compares various energy balance models (EBMs) through a series of numerical experiments. The objective is to establish rigorous protocols that enable the identification of intrinsic differences among EBMs that could lead to model-dependent results for past, current, and future EBM studies. Such efforts also provide the community with an EBM ensemble average and standard deviation, rather than a single model prediction, on benchmark cases typically used by EBM studies. These experiments include Earth-like planets at different obliquity, instellation, and CO$_2$ abundance. Here we update the v1.0 protocol \citep{Deitrick2023} to accommodate the requirements of previously untested community models. In particular, we expand the range of CO$_2$ abundances for Experiment 4 to ensure any code will capture both snowball and ice-free end states. Additionally, participants are now required to report two ice edge latitudes per hemisphere to fully distinguish all climate states (snowball, ice caps, ice belts, and ice-free). The outputs described in FILLET protocol version 1.0 have also now been revised  to include the maximum and minimum ice extent, in latitude, for each hemisphere, as well as the diffusion coefficient and outgoing longwave radiative flux.
\end{abstract}

\section{Introduction}

Planetary climate models offer an opportunity to characterize the surface habitability of planets and moons, but results and predictions can vary substantially between models. Since instrument and mission development plans can be guided by climate simulations, the Climates Using Interactive Suites of Intercomparisons Nested for Exoplanet Studies (CUISINES) meta-framework was formed to develop explicit methodologies, or exoplanet intercomparison project (exo-MIP) protocols, to compare results between community models \citep{Sohl2024}. After the model community generates their outcomes of the protocol, separate results papers are published (i.e. \cite{turbet2021, sergeev2021, Fauchez2021} for the THAI exo-MIP) that analyze the differences between models and predictions. 

As part of CUISINES, the Functionality of Ice Line Latitudinal EBM Tenacity (FILLET) exo-MIP considers energy balance models (EBMs), which are relatively simple one- or two-dimensional climate models that can calculate the spatial distribution of temperature and ice coverage on the surfaces of Earth-like planets \citep[see, e.g.,][]{northcoakley1979,williamskasting97,north2017,okuya2019,biasiotti2022,Ramirez24}. In 2023, the first FILLET protocol was established \citep[v1.0; ][]{Deitrick2023} and numerous teams submitted their results. However, some codes could not fully participate because some v1.0 parameter ranges were too restrictive. We therefore present v1.1 below, which expands several parameter ranges to enable the comparison of more EBMs and also updates the required model output. This revision enables robust model intercomparisons that will be published in the forthcoming results paper.

The following section primarily describes the revised values of tabular data in v1.0. For detailed discussions of the motivation, requirements, and descriptions of the v1.0 Benchmarks and Experiments, the reader is referred to \cite{Deitrick2023}. To participate in the FILLET exo-MIP, please visit the project website\footnote{\url{https://nexss.info/cuisines/fillet-mip/}} and follow the instructions and/or contact the current FILLET project leader, or ``chef''.

\section{The v1.1 Protocol}

This section provides an overview of the changes to the v1.0 protocol, including the required model outputs. Benchmark 1 (both ``tuned'' and ``untuned'') remains unchanged as it is a reproduction of the pre-industrial Earth, without any model parameter restrictions. The issues with the v1.0 protocol arose when some codes that employ different climate parameterizations could not be pushed into the climate regimes that the experiments required within the allowed parameter ranges. These models and codes were not wrong or buggy, they were just sufficiently different from the codes used to define the v1.0 protocol that Benchmarks 2--3 and Experiments 1--4 must be revised.

Table \ref{tab:Params1} provides the new values for the updated Benchmarks and Experiments. This table replaces Table 3 from \cite{Deitrick2023}. In particular, note that the CO$_2$ abundance range for Experiment 4 has increased from [50,5050] ppm to [1,$10^5$] ppm. This extension should ensure that all EBMs capture both snowball and ice-free climate states. Note that, as in v1.0, this table does not include Benchmark 1 as there are no guidelines for it other than reproducing the Earth's surface temperature of 288 K.

\begin{table*}[]
\caption{Protocol parameters (experiment specific). Square brackets indicate spacing of varied parameters.}
\centering
\begin{tabular}{lcccc}
\hline
\hline
 & Instellation & CO$_2$ abundance & Semi-major axis & Obliquity \\
 & $S$ ($S_{\oplus}$)$^\dagger$ & $X_{\text{CO}_2}$ (ppm)$^\ddagger$ & $a$ (au)$^\star$ & $\varepsilon$ ($^{\circ}$) \\
\hline
{\bf Benchmark 2:}\\
Un-tuned, low obliquity & 1 & 280 & 1 & 23.5 \\
\\
{\bf Benchmark 3:}\\
Un-tuned, high obliquity & 1 & 280 & 1 & 60 \\
\\
{\bf Experiment 1:}\\
G-dwarf warm start & 0.8-1.25$^{\star\star}$ [0.025] & 280 & 1 & 0-90 [10$^{\circ}$] \\
 \\
{\bf Experiment 1a:}\\
G-dwarf warm start, & Set by $a^{\star\star\star}$ & 280 & 0.875-1.1$^{\star\star}$ [0.0125] & 0-90 [10$^{\circ}$]  \\
orbit variation & & &  & \\
\\
{\bf Experiment 2:}\\
G-dwarf cold start & 1.05-1.5$^{\star\star}$ [0.025] & 280 & 1 & 0-90  [10$^{\circ}$] \\
\\
{\bf Experiment 2a:}\\
G-dwarf cold start, & Set by $a^{\star\star\star}$ & 280 & 0.8-0.975$^{\star\star}$ [0.0125]  & 0-90 [10$^{\circ}$]  \\
orbit variation & & & &  \\
\\
{\bf Experiment 3:}\\
Bifurcation diagram,  & 0.8-1.5$^{\star\star}$ [0.0125] & 280 & 1 & 23.5 \\
varying instellation \\
\\
{\bf Experiment 4:}\\
Bifurcation diagram,  & 1 & 1-100,000$^{\star\star}$ [50 pts$^{\star\star\star\star}$] & 1 & 23.5  \\
varying $X_{\text{CO}_2}$ \\
\\
\hline
\end{tabular}\\
\vspace{0.5em}
$^\dagger$ $S_{\oplus} = 1361$ W m$^{-2}$.\\
$^\ddagger$ CO$_2$ volume mixing ratio in parts-per-million in a 1 bar, N$_2$-dominated atmosphere.\\
$^\star$ 1 au = $1.495978707\times10^{11}$ m.\\
$^{\star\star}$ Ranges should be extended if necessary to capture both snowball and ice-free states.\\
$^{\star\star\star}$ Use the definition $S(a) = S_{\oplus}/a^2$ to scale the instellation. \\
$^{\star\star\star\star}$ For Experiment 4, the grid is 50 points with logarithmic spacing. 
\label{tab:Params1}
\end{table*}

In addition to revised parameter ranges, v1.1 updates the required outputs for the Benchmarks and Experiments. The updated output products are provided in Table \ref{tab:Outputs}, which replaces Table 5 in \cite{Deitrick2023}. The v1.1 protocol makes two key changes from v1.0, which are described below. The global output template in the FILLET website has also been updated to include the output requirements for v1.1.

The first update increases the number of reported ice edge latitudes per hemisphere from 1 to 2, which ensures discrimination between snowball, ice caps (ice sheets that encompass one pole but do not extend to the other), ice belts (ice sheets that do not cover either pole), and ice free states. The v1.0 protocol could lead to the misidentification of ice caps and/or ice belts in some cases because without two ice lines for each surface type, it can be unclear if land and/or sea ice sheets extend to the equator or to the pole. The outputs described in Table \ref{tab:Outputs} now include the minimum and maximum ice edge latitude for each hemisphere, the diffusion coefficient, and the outgoing longwave radiation (OLR). Participants should therefore identify the latitudinal extent of land/sea ice sheets to classify the climate state as either ice-free, polar caps, ice belt, or snowball.

A second issue with v1.0 arises for models that contain separate grid points for ocean and land, such as the Bitz family models \citep[e.g.,][]{shields2013,deitrick2018,rushby2019,barnes2020, Venkatesan2025, Venkatesan_EBM_2025}. Depending on the modeling choices, especially surface albedos, the land and ocean components can have different ice edge latitudes, resulting in misclassifications of the climate state. As a consequence, v1.1 requires users of this type of model to provide 4 ice edge values per hemisphere: 2 for the land and 2 for the sea. For ice belts that span the equator, the northern minimum and southern maximum should be set to 0 even though there is no ice edge there.

\begin{table*}[]
\caption{Outputs}
\centering
\begin{tabular}{lcc}
\hline
\hline
Output type & Columns (units) & Output header name\\
\hline
{\em Latitudinally-varying outputs:}
 & Latitude ($^{\circ}$) & Lat\\
 & Annually-averaged surface temperature (K) & Tsurf\\
 & Annually-averaged surface albedo & Asurf\\
 & Annually-averaged TOA albedo & ATOA \\
 & Annually-averaged outgoing longwave radiation & OLR\\
\\
{\em Global outputs:}
 & Instellation ($S_{\oplus}$)$^{\dagger}$ & Inst\\
 & Obliquity ($^{\circ}$) & Obl\\
 & $X_{\text{CO}_2}$ (ppm)$^\ddagger$ & XCO2 \\
 & Global mean surface temperature (annual average) (K) & Tglob\\
 & Maximum latitude of ice  on land in the Northern hemisphere ($^{\circ}$) & IceLineNMaxLand\\
 & Minimum latitude of ice on land in the Northern hemisphere ($^{\circ}$) & IceLineNMinLand\\
 & Maximum latitude of ice on the sea in the Northern hemisphere ($^{\circ}$) & IceLineNMaxSea\\
 & Minimum latitude of ice on the sea in the Northern hemisphere ($^{\circ}$) & IceLineNMinSea\\
 & Maximum latitude of ice on land in the Southern hemisphere ($^{\circ}$) & IceLineSMaxLand\\
 & Minimum latitude of ice on land in the Southern hemisphere ($^{\circ}$) & IceLineSMinLand\\
 & Maximum latitude of ice on sea in the Southern hemisphere ($^{\circ}$) & IceLineSMaxSea\\
 & Minimum latitude of ice on sea in the Southern hemisphere ($^{\circ}$) & IceLineSMinSea\\
 & Heat diffusion coefficient (W m$^{-2}$ K$^{-1}$) & Diff \\
 & Global mean outgoing longwave radiation (W m$^{-2}$) & OLRglob \\
\hline
\end{tabular}\\
$^{\dagger}$ $S_{\oplus} = 1361$ W m$^{-2}$. 
$^\ddagger$ CO$_2$ volume mixing ratio in parts-per-million. \\
\label{tab:Outputs}
\end{table*}

The inputs described in this protocol and the scripts necessary for the analysis of the data and the production of plots for publications will be made available on the FILLET GitHub repository\footnote{\url{https://github.com/projectcuisines/fillet}} and Zenodo\footnote{\url{https://doi.org/10.5281/zenodo.7542262}}. The inputs are available immediately, while the scripts to reproduce the results will be made publicly available upon the publication of the results. Model outputs will also be made permanently available in the NASA EMAC   repository\footnote{\url{https://ckan.emac.gsfc.nasa.gov/organization/cuisines-fillet}}.
\section{Conclusions}

We have presented a revision to the FILLET v1.0 protocol \citep{Deitrick_2023} that permits more codes to participate in the exo-MIP. Our updates have relaxed parameter ranges to nearly their physical limits, so v1.1 should not need further revision. Participants should use the parameter ranges in Tables \ref{tab:Params1}--\ref{tab:Outputs} when performing the Benchmark and Experiment simulations. 

The next step for FILLET will be the presentation of the contributed EBM results to identify commonalities and differences between the participating models in follow-up papers to be published in the CUISINES focus issue\footnote{\url{https://iopscience.iop.org/collections/2632-3338_CUISINES}}. Ultimately, the FILLET exo-MIP will provide a community ensemble average and standard deviation of EBM predictions for standard test cases, similar to the Coupled Model Intercomparison Project (CMIP, \cite{Eyring_2016}) for Earth GCMs, which can avoid many of the dependencies, defects, and biases that can arise from using a single model in isolation. Such an endeavor will make EBM predictions more robust, which would help guide mission planning and perhaps even contribute to the discovery of life beyond the Solar System. 

\vspace{1cm}

\noindent{\bf Acknowledgments}\\
FILLET belongs to the CUISINES meta-framework, a Nexus for Exoplanet System Science (NExSS) science working group\footnote{\url{https://nexss.info/cuisines/}}. R.B. and T.J.F. are supported by the GSFC Sellers Exoplanet Environments Collaboration (SEEC), which is funded in part by the NASA Planetary Science Divisions Internal Scientist Funding Model. R.B. is also funded by the ROCKE-3D consortium through NASA award No. 80NSSC24K0856. Financial support to R.D. was provided by the Natural Sciences and Engineering Research Council of Canada (NSERC; Discovery Grant RGPIN-2018-05929), the Canadian Space Agency (Grant 18FAVICB21), and the European Research Council (ERC; Consolidator Grant 771620). P.S. acknowledges the Italian National Institute of Astrophysics (INAF) for funding his work under the INAF mini-grant titled \textit{Radiative models for the paleoatmospheres of Mars, Earth and Venus} (F.O. 1.05.121.04.03). J.H.M. acknowledges funding from the NASA Exobiology program under award 80NSSC22K1632. V.V. acknowledges support from the NASA FINESST Fellowship under award number 80NSSC21K1852. Data and scripts for FILLET are available at DOI: \dataset[10.5281/zenodo.7563242]{\doi{10.5281/zenodo.7563242}}.

\bibliographystyle{aasjournal}
\bibliography{main,version1_1}
\end{document}